# Vector Soliton Breathing Dynamics

Zhiwei Huang[1](黄志玮), Sergey Sergeyev[2*], Qing Wang(王清)[2], Hani Kbashi[2], Dmitrii Stoliarov[2], Qianqian Huang[1](黄千千), Yuze Dai[3](代雨泽), Zhijun Yan[3](闫志君), and Chengbo Mou[1*](牟成博)

[1]*Key Laboratory of Specialty Fiber Optics and Optical Access Networks, Joint International Research Laboratory of Specialty Fiber Optics and Advanced Communications, Shanghai Institute for Advanced Communication and Data Science, Shanghai University, 200444, China*

[2]*Aston Institute of Photonic Technologies (AIPT), Aston University, Aston Triangle, Birmingham, B4 7ET, UK*

[3]*The School of Optical and Electronic Information and NGIA, Huazhong University of Science and Technology, Wuhan,430074, China*

*Corresponding authors e-mail addresses: s.sergeyev@aston.ac.uk, mouc1@shu.edu.cn*

## Abstract

**Mode-locked lasers play the role of the ideal testbeds for studying self-coherent structures - dissipative solitons – with stable spatiotemporal profiles supported by the balance between dispersion and nonlinearity. However, under some conditions, the profile and energy of the solitons can oscillate (breath) periodically. Unlike the previous studies of different scalar mechanisms of breathers' emergence, we demonstrate experimentally and theoretically a new vector mechanism where breathing of the output power is a result of the heteroclinic dynamics. The heteroclinic orbit is a trajectory periodically evolving nearby the neighborhood of one of the orthogonal states of polarization with further switching to and evolving nearby the other state of polarization. The dwelling time for the trajectory near each orthogonal state is determined by the cavity anisotropy adjustable with the help of the polarization controller for the pump wave. The obtained results on tunability of breathing dynamics enables unlocking novel techniques of the flexible control and manipulation of the polarization dynamics of dissipative solitons.**

Keywords: breather, polarization instability, fiber laser, mode-locking, polarization

During the last three decades, the study of flocking birds, supramolecular complexes, neurons in the cortex, modes synchronization in lasers, and telecom and sensing networks are focused on revealing how the interactions between individual system components produce large-scale collective patterns [1-4]. However, in the practical context, targeting the collective patterns under demand is challenging due to the limited ability to conduct experiments on manipulating engineering and biological networks' structure [1-4]. The short pulse duration of hundred femtoseconds and repetition rates of tens–hundreds of megahertz make multimode mode-locked lasers (MLLs) suitable testbeds for studying the synchronization-driven self-organization under controllable laboratory conditions and short time scale of seconds [5-22]. The synchronization leads through short-range (covalent) and long-range (non-covalent) weak pulses' interactions toward swarming pulses into different soliton supramolecules [7-19].

Manipulating the collective patterns in MLL is based on controlling short-range and long-range interactions among dissipative solitons (DSs) -localized waves arising from a balance of dissipative and dispersive effects [5, 6]. Short-range interaction through the overlapping of solitons tails or soliton-dispersive wave interaction results in bound states (BS) solitons which, in analogy to biochemical and biological supramolecules, are formed by strong covalent bonds, frequently referred to as soliton molecules, soliton macromolecules, or soliton crystals [7-9]. The strong, short-range interactions lead to the narrow spacing of the few pulse widths and locked phase differences between adjacent solitons, resulting in the highly challenging real-time characterization of their detailed temporal structure [7-9]. Long-range interactions can be driven by Casimir-like [10, 11], optoacoustic [9,12-14], polarization instabilities [15,16], and soliton-dispersive wave interaction [13-15] and lead to the formation of the soliton structures in the form of separate multi-pulsing, harmonic mode-locking, soliton rain, and the breathers [8-21]. In the case of breathers, the profile and energy of DSs can oscillate (breath) with a period ranging from hundreds to ten thousand roundtrips [18-21]. It has been recently found that the breather emergence in anomalous dispersion mode-locked fiber lasers can be caused by modulation instability (MI) [18]. Unlike this, the breathers in the normal dispersion regime can arise because of Hopf bifurcation [19], polarization properties of the ML laser [20], and subharmonic entrainment (SHE) [21]. Also, by accounting for the vector nature of DSs, it has been recently demonstrated experimentally for Er-doped laser mode-locked with carbon nanotubes that adjust the coupling between two orthogonally polarized polarization states (SOPs) leads to polarization instabilities resulting in the soliton breathing [22]. In this Letter, we explore experimentally and theoretically a universal nature of the breather's emergence caused by polarization instabilities. Unlike our previous study, we use Er-doped fiber laser mode-locked based on nonlinear polarization rotation and normal dispersion along with a vector model. For the first time, we reveal experimentally and theoretically that similar to the previous case, the breather emergence is caused by the desynchronization of the orthogonally polarized modes. In contrast to the previous scalar forms of breathers with periods of 10-100 roundtrips (RTs) [18-21], the observed breathing dynamics has much more extended periods of 10000 RTs.

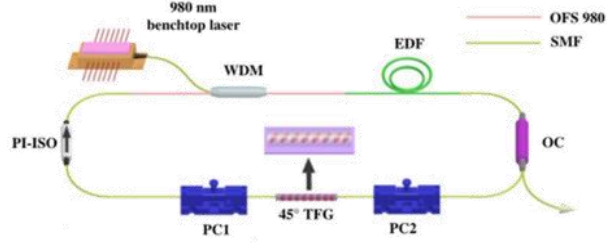

FIG. 1 (color online). Schematic setup of the NPR mode-locked fiber laser. Erbium doped fiber laser. EDF: erbium-doped fiber; LD: 980 nm laser diode for pump; PC1 and PC2: polarization controllers, PI-ISO: polarization insensitive isolator; 45°-TFG- 45°-tilted fiber grating based polarizer; WDM: wavelength division multiplexer (WDM), OC: 91:9 % output coupler.

The configuration of NPR mode-locked fiber laser is shown in the Fig.1 (details are found in Supplementary Material [23]). By using the fast photodetector and oscilloscope we record the dynamics waveforms. By utilizing commercial polarimeter (THORLABS, IPM5300) with 1 μs resolution, we can observe the evolution of the polarization attractors at the Poincare sphere in terms of the normalized Stoke parameters $s_1$, $s_2$ and $s_3$, the power for the orthogonal x- and y- polarization components ($I_x$, $I_y$) and the total power $S_0$, the phase difference $\Delta\varphi$ and degree of polarization DOP.

$$S_0 = I_x + I_y, \ S_1 = I_x - I_y, \ S_2 = 2\sqrt{I_x I_y} \cos \Delta \varphi, \ S_3 = 2\sqrt{I_x I_y} \sin \Delta \varphi,$$
$$s_i = S_i/\sqrt{S_1^2 + S_2^2 + S_3^2}, \quad DOP = \sqrt{S_1^2 + S_2^2 + S_3^2}/S_0, (i = 1,2,3). \quad (1)$$

the fast and slow dynamics from breather to Q-switched and continuous wave (CW) mode-locking are shown in Fig.2 – Fig. 4 for different pump powers, e. g. $I_p$=500 mW (Fig.2 a-d and Fig. 3 a-d), $I_p$=260 mW (Fig. 4, a-d) and adjusted in-cavity polarization controllers.

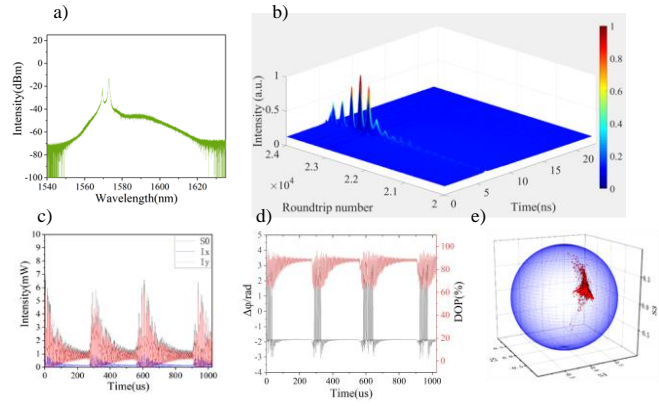

FIG. 2 (color online) Breather dynamics: a) optical spectrum, b) oscilloscope trace; c) slow polarimeter trace of the output powers of the orthogonally polarized components $I_x$ (blue) $I_y$ (red) and total $I=I_x+I_y$ (black); d) slow polarimeter trace of the phase difference $\Delta\varphi$ (black) and DOP (red); e) slow polarimeter trace of the trajectories on the Poincaré sphere. Pump power $I_p$=500 mW.

As follows from the recent publication [20] and Fig. 2(a), the optical spectrum has two maxima that reflect the breathing spectral dynamics. The internal structure of a single breather shown in Fig.2 (b) reveal the breather width of the 200 round trips (RTs) with the breather period of 10000 RTs (recorded with 1 μs resolution by polarimeter) shown in Fig 2 (c). Fig. 2 (d) indicates that the breather's power spikes emergence and disappearance is related to the periodic phase difference slip in π and DOP hops and so to the longitudinal and orthogonal polarization modes synchronization and desynchronization. As shown in Fig. 2 (e), the trajectories in the Poincaré sphere also take the form of hops from the localized SOPs. So, figures 2(a-e) support the conclusion that the vector soliton breathing dynamics includes alteration of the longitudinal and polarization modes synchronization and desynchronization [24].

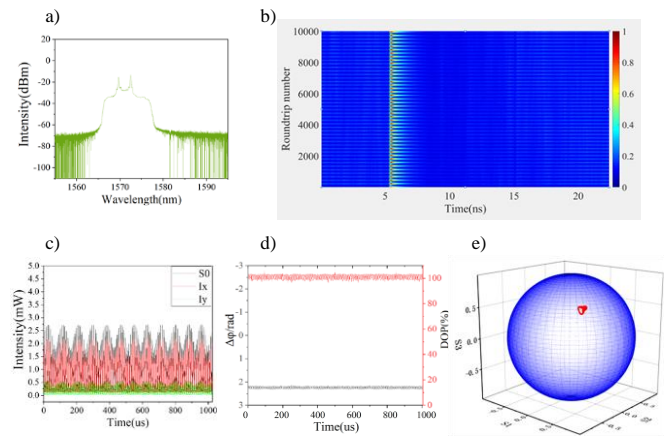

FIG. 3 (color online) Breathing dynamics in the form of Q-switched mode-locking: a) optical spectrum, b) temporal trace; c) slow polarimetric trace of orthogonally polarized components $I_x$ (blue) $I_y$ (red) and total $I=I_x+I_y$

(black); d) phase difference $\Delta\varphi$ (black) and DOP (red); e) slow polarimetric trajectories on Poincaré sphere. Pump power $I_p$=500 mW.

Similar to previous case shown in Fig. 2(a), the optical spectrum in Fig. 3(a) also exhibits two maxima [20]. However, the breathing dynamics take the form of two-scale oscillations (Q-switch mode-locking) with the periods of 200 RTs (Fig. 3(b)) and 5000 RTs (Fig. 3(c)). As follows form Fig. 3(d), the dynamics is caused by the phase difference entrainment (oscillations). The DOP in Fig. 3(d) is of 90 % indicating that the dynamic is slow at the time scale of 1 μs – 1 ms from which SOP of each single pulse could be resolved. As shown in Fig. 3(e), the trajectory in the Poincaré sphere is a cycle and so the vector soliton breathing dynamics takes the form of phase difference entrainment synchronization scenario [25].

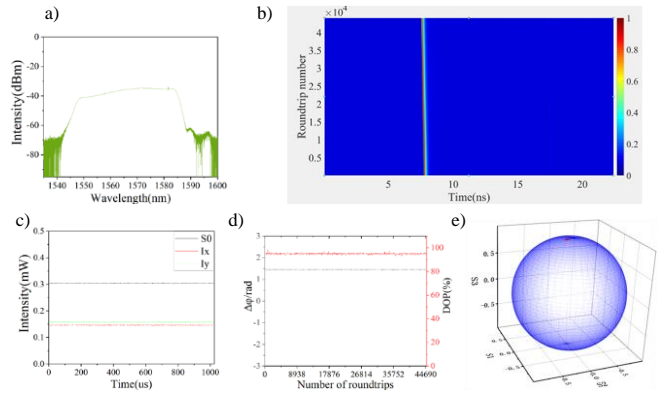

FIG. 4 (color online) CW mode-locking: a) optical spectrum, b) oscilloscope trace; c) slow polarimeter trace of the output powers of the orthogonally polarized components $I_x$ (blue) $I_y$ (red) and total $I=I_x+I_y$ (black); d) slow polarimeter trace of the phase difference Δφ (black) and DOP (red); e) slow polarimeter trace of the trajectories on the Poincaré sphere. Pump power $I_p$=260 mW.

The continuous wave (CW) mode-locking case is shown in Figs. 4 (a-e). The wide optical spectrum typical for the normal dispersion operation is shown in Fig. 4(a) [20]. The pulse train has a stable amplitude with the small variation of the peak power at the fast and slow time scales as shown in Fig.4(b, c). The CW mode-locking corresponds to the case of fixed phase difference and SOP locking with high DOP above 90% (Fig. 4 (d, e)).

Approach which was previously used for theoretical characterization of NPR-based mode-locking features introduction of function Ψ which is related to amplitudes of the polarization components $E_x$ and $E_y$ as follows [25]:

$$E_x = \Psi \cdot cos(\theta), E_y = \Psi \cdot sin(\theta). \qquad (2)$$

Here $\theta$ is the angle between the polarization plane of the polarizer and slow axis [26]. The complex amplitudes of the orthogonally polarized SOPs $E_x$ and $E_y$ can be also presented as follows:

$$E_x = |E_x|\exp(i \cdot \varphi_x), E_y = |E_y|\exp(i \cdot \varphi_y). \tag{3}$$

As follows from Eqs. (2) and (3), the phase difference $\Delta\varphi = \varphi_y - \varphi_x \equiv 0$, and $tg(\theta) = |E_y|/|E_x|$. However, as follows from Figs. 2(d) and 3(d), the phase difference $\Delta\varphi$ is oscillating that makes Eq. (2) not acceptable for the modelling of the polarization dynamics of the breathing solitons.

To clarify contribution of the polarization properties of the cavity to the breathing dynamics, we used the vector model of mode-locked fiber laser accounting for slow polarization dynamics of the lasing field in the form of the Stokes vector $\mathbf{S} = (S_0, S_1, S_2, S_3)^T$ and active medium in terms of the functions $f_1, f_2$, and $f_3$, accounting for the vector nature of the interaction of the lasing field with an active medium [27-29] (details are found in Supplementary Material [23]). To mimic contribution of two polarization controllers and polarizer shown in Fig. 1, we account for the linear $\beta_L$ and circular $\beta_C$ birefringence ($\beta_{L\,(C)} = 2\pi L/L_{bL(bC)}$, $L_{bL(bC)}$ is the linear (circular) birefringence beat length and $L$ is the cavity length) and the pump anisotropy $\xi = (1 - \delta^2)/(1 + \delta^2)$, where $\delta$ is the ellipticity of the pump wave [27-29]. Given that the dynamics in the model was averaged over the pulse shape, the breathing soliton dynamics corresponds to the phase difference desynchronization, the slow complex oscillations (Q-switched mode locking), and finally the steady state operation (CW mode locking).

To identify the range parameters, e. g. the normalized pump power $I_p$, pump anisotropy $\xi$, and the linear birefringence $\beta_L$, mapping the polarization modes desynchronization, the phase difference entrainment and locking, we linearized the Eqs. (SE 1) in the vicinity of the steady state solution ($S_0 \neq 0, S_1 = S_2 = 0, S_3 = \pm S_0$) and found numerically eigenvalues for the parameters quite close to the experimental ones. As a result, eigenvalues $\lambda$ and saddle index $\nu$ take the form [27]:

$$\lambda_0 = 0, \quad \lambda_{1,2} = -\gamma_1 \pm i\omega_1, \quad \lambda_{3,4} = -\gamma_2 \pm i\omega_2, \quad \lambda_{5,6} = \rho \pm i\omega_3, \quad \nu = \left|\frac{\gamma_1}{\rho}\right|$$

$$(\omega_{1,2,3} \neq 0, \ \rho, \gamma_1, \gamma_2 > 0, \ \gamma_1 > \gamma_2). \tag{4}$$

Given that eigenvalues for steady states ($S_0 \neq 0, S_1 = S_2 = 0, S_3 = S_0$) and ($S_0 \neq 0, S_1 = S_2 = 0, S_3 = -S_0$) are equal, condition for the orthogonal SOPs desynchronization takes the form $\nu < 1$ [27, 28, 30, 31].

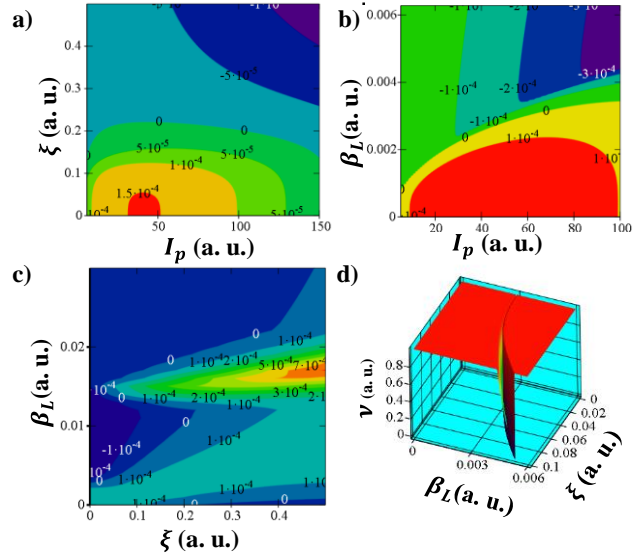

FIG. 5 (color online) (a)-(c) The maps of self-pulsing scenarios (Re{$\lambda_5$}>0, Im{$\lambda_5$}≠0) and (c) saddle index ($\nu$=| Re{$\lambda_1$}/ Re{$\lambda_5$}| <1) as a function of the pump power $I_p$, linear birefringence $\beta_L$ and the anisotropy of pump $\xi = (1 - \delta^2)/(1 + \delta^2)$ in the vicinity of the steady state solution ($S_0$≠0, $S_1$=$S_2$=0, $S_3$=±$S_0$). Parameters can be found in Supplementary Material [23].

As follows from Eqs. (4), oscillatory behavior is emerging when $\rho > 0$. The corresponding contour plots are shown in Fig. 5 (a-c). The saddle index $\nu$ as a function of the anisotropy of the pump $\xi$ and birefringence strength $\beta_L$ is shown in Fig. 5 (d). The area inside surface shown in Fig. 5 (d) and the surface $\nu = 1$ is related of the area with breathing oscillations [27]. As follows from Figs. 5 (a) - (d) oscillatory regimes exists for the wide range of the pump power and very narrow range of the pump power anisotropy and the in-cavity birefringence strength. Tunability of the dynamics with tuning the ellipticity of the pump wave is shown in Figs. 6 - 9.

As follows from Fig.6, the weak linear birefringence and low anisotropy of the pump wave within the range of parameters corresponding the case $\nu < 1$ in Fig. 5 (d) can lead to the breathing regime. The breather polarization dynamics shown in Figs. 6 (a)-(c) is quite close to the dynamics shown in Fig. 2 (c) and (d) in the context of the shape and the period (approx. 10000 RTs) of the breathing and the phase difference slips in $\pi$ radian. The main difference is in the Poincaré sphere trajectories that can be caused by the simplified matter of the model that doesn't take into account the dual-wavelength lasing shown in Fig. 2 (a) and slow polarimeter operation with sampling of 50 RTs.

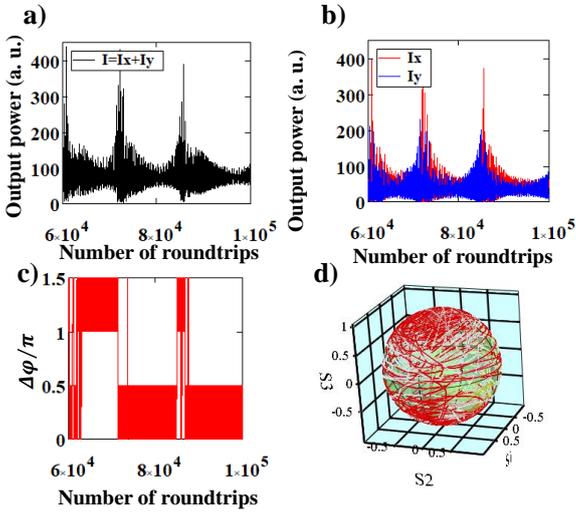

FIG. 6 (color online) The main parameters: $\xi = 0.11$, $\beta_L = 2\pi \cdot 0.001, \beta_c = 0, I_p = 55$ (a) Breathing polarization dynamics in the form of complex oscillations of the output power total power $I = I_x + I_y$; (b) the powers of the polarization components $I_x$ and $I_y$; (c) the phase difference $\Delta\varphi$ and (d) SOP trajectories in Poincaré sphere.

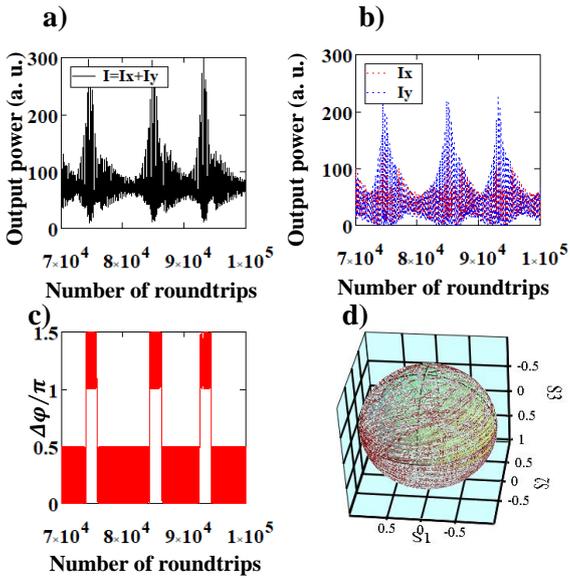

FIG. 7 (color online) The main parameters: $\xi = 0.05, \beta_L = 0, \beta_c = 2\pi \cdot 0.001, I_p = 55$. Breathing polarization dynamics in the form of complex oscillations of the output power total power $I = I_x + I_y$ (a); the powers of the polarization components $I_x$ and $I_y$ (b); the phase difference $\Delta\varphi$ (c); and trajectories in Poincaré sphere (d).

Though breathing regimes were not mapped in terms of the circular birefringence, the dynamics can also emerge for the weak circular birefringence and low anisotropy of the pump wave as shown in Fig. 7 (a-d). It was recently found that the weak coupling of the polarization components $x$ and $y$ for the case of weak birefringence and low anisotropy leads to the phase difference desynchronization [25, 27-29], that in our case corresponds to the breathing polarization dynamics.

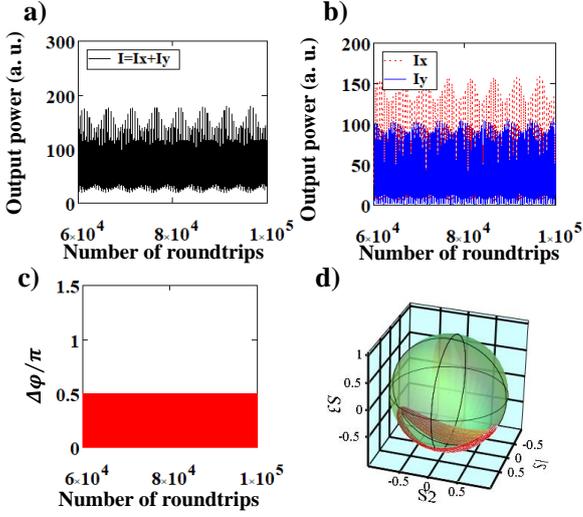

FIG. 8 (color online) The main parameters: $\xi = 0.22, \beta_L = 0, \beta_c = 2\pi \cdot 0.001, I_p = 55$. Complex polarization dynamics in the form of complex oscillations (Q-switched mode-locking) of the output power total power $I = I_x + I_y$ (a); the powers of the polarization components $I_x$ and $I_y$ (b); the phase difference $\Delta\varphi$ (c); and trajectories in Poincaré sphere (d).

Increased the pump anisotropy from $\xi = 0.11$ to $\xi = 0.22$ leads to the enhanced coupling of the polarization components $x$ and $y$ and so to the periodic oscillations of the total's (Fig. 8 (a)) and polarization components' powers (Fig. 8(b)), along with the periodic oscillations of the phase difference (Fig. 8 (c)). The polarization dynamics is quite similar to that shown in Fig. 3 (c, d). The main difference is in the trajectories in Poincaré sphere that can be caused by the doesn't take into account the dual-wavelength lasing shown in Fig. 3 (a) and slow polarimeter operation with sampling of 50 RTs.

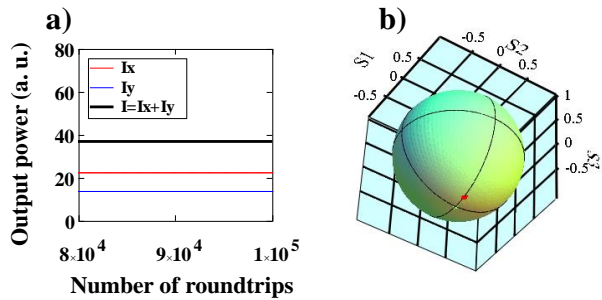

FIG. 9 (color online) Steady state (polarization-locked regime) in the form of the constant output powers (a) and locked state of polarization (b). The main parameters: $\xi = 0.22, \beta_L = 0, \beta_c = 0, I_p = 29$.

As follows from the Fig. 5(a) and our previous study [24, 27], the combination reduced pump power from $I_p = 55$ to $I_p = 29$ and high pump anisotropy $\xi = 0.22$ leads to the increase coupling between $x$ and $y$ components and so to the steady-state operation (CW or polarization-locked regime) shown in Fig. 9 (a, b) which corresponds to the experimentally observed case shown in Fig. 4 (b-e).

Though the vector model demonstrates just qualitative correspondence to the experimental data, it has applicability for interpreting results on the breather dynamics in the case of CNT-based mode-locking [22] and so reveals the universal nature of the vector breathing dynamics.

In conclusion, we reveal experimentally and theoretically a new mechanism of the breather's dynamics emergence caused by the phase difference desynchronization of two orthogonally polarized SOP in Er-doped fiber laser mode-locked by NPR. The breathers appears when we adjust the coupling strength between the orthogonal SOPs by tuning the pump power and polarization controllers. The developed presentation of the breather's emergence as desynchronization events in the system of coupled oscillators show great potential for mapping conditions for the complex dynamics emergence and so for developing techniques for flexible control of the laser's vector dynamics. The technique can be based on the main result related to demonstrating that the mode-locked fiber laser with low cavity anisotropy presents a heteroclinic system where the laser's eigenstates – orthogonal states of polarization - are quasi-equilibrium points. As follows from the Shil'nikov theorem [30, 31], the heteroclinic system produces a large number of attractors which are located near the heteroclinic orbit connecting the quasi-equilibrium states. We have outlined in previous publication that injected an optical signal with the dynamically evolving state of polarization can enable flexible control and manipulation of the heteroclinic dynamics towards different stable dynamically evolving SOPs [32].

We acknowledge support from the National Natural Science Foundation of China (61975107,62075071, 61605107), the '111' project (D20031), UK EPSRC (EP/W002868/1), Horizon 2020 ETN MEFISTA (861152) and EID MOCCA (814147).

[29] Supplementary

*Experimental set-up.*

A 1.48 m erbium-doped fiber (EDF) with a normal group velocity dispersion (GVD) of +66.1 ps$^2$/km is used in this cavity. Besides, the cavity also contains 0.9 m of OFS980 fiber with a normal GVD of + 4.5 ps$^2$/km and 2.34 m of a single mode fiber (SMF) with an anomalous GVD of -21.67 ps$^2$/km. The total length of the cavity is 4.72 m, corresponding to the fundamental frequency of 44.18 MHz and the net dispersion of the cavity is + 0.046 ps$^2$ and so the laser operates in the normal dispersion. The pump light is coupled to the laser cavity through a 980 / 1550 nm wavelength division multiplexer (WDM). A 91:9 coupler is used to direct out 9% of the pulse energy outside the cavity. The polarization-insensitive isolator (PI-ISO) in the cavity makes the unidirectional transmission of the pulse train. A fiber polarizer and the two polarization controllers (PCs) are used to support nonlinear polarization rotation (NPR) mechanism for passive mode-locking.

*Vector model of Er-doped fiber laser*

Evolution of the laser SOPs and population of the first excited level in Er$^{3+}$ doped active medium was modeled using the following equations derived from the vector theory developed by Sergeyev and co-workers [26-28]:

$$\frac{dS_0}{dt} = \left(\frac{2\alpha_1 f_1}{1+\Delta^2} - 2\alpha_2\right) S_0 + \frac{2\alpha_1 f_2}{1+\Delta^2} S_1 + \frac{2\alpha_1 f_3}{1+\Delta^2} S_2,$$

$$\frac{dS_1}{dt} = \gamma S_2 S_3 + \frac{2\alpha_1 f_2}{1+\Delta^2} S_0 + \left(\frac{2\alpha_1 f_1}{1+\Delta^2} - 2\alpha_2\right) S_1 - \beta_c S_2 - \frac{2\alpha_1 f_3 \Delta}{1+\Delta^2} S_3,$$

$$\frac{dS_2}{dt} = -\gamma S_1 S_3 + \frac{2\alpha_1 f_3}{1+\Delta^2} S_0 + \beta_c S_1 + \left(\frac{2\alpha_1 f_1}{1+\Delta^2} - 2\alpha_2\right) S_2 + \frac{2\alpha_1 f_2 \Delta}{1+\Delta^2} S_3,$$

$$\frac{dS_3}{dt} = \frac{2\alpha_1 \Delta f_3}{1+\Delta^2} S_1 - \frac{2\alpha_1 \Delta f_2}{1+\Delta^2} S_2 + \left(\frac{2\alpha_1 f_1}{1+\Delta^2} - 2\alpha_2\right) S_3,$$

$$\frac{df_1}{dt} = \varepsilon\left[\frac{(\chi_s - 1)I_p}{2} - 1 - \left(1 + \frac{I_p \chi_p}{2} + d_1 S_0\right) f_1 - \left(d_1 S_1 + \frac{I_p \chi_p}{2}\frac{(1-\delta^2)}{(1+\delta^2)}\right) f_2 - d_1 S_2 f_3\right],$$

$$\frac{df_2}{dt} = \varepsilon\left[\frac{(1-\delta^2)}{(1+\delta^2)}\frac{I_p(\chi_s - 1)}{4} - \left(\frac{I_p \chi_p}{2} + 1 + d_1 S_0\right) f_2 - \left(\frac{(1-\delta^2)}{(1+\delta^2)}\frac{I_p \chi_p}{2} + d_1 S_1\right)\frac{f_1}{2}\right],$$

$$\frac{df_3}{dt} = -\varepsilon\left[\frac{d_1 S_2 f_1}{2} + \left(\frac{I_p \chi_p}{2} + 1 + d_1 S_0\right) f_3\right].$$

(SE 1)

Here time and length are normalized to the round trip and cavity length, respectively; $S_i$ ($i=0,1,2,3$) are the Stokes parameters; $S_0$ and $I_p$ are the output lasing power and the pump normalized to the corresponding saturation powers; $\beta_{L\,(C)} = 2\pi L/L_{bL(bC)}$ is the linear (circular) birefringence strength normalized to the cavity length $L$, $L_{bL(bC)}$ is the linear (circular) birefringence beat length; $\alpha_1$ is the total absorption of erbium ions at the lasing wavelength, $\alpha_2$ is the total insertion losses in the cavity; $\delta$ is the ellipticity of the pump wave, $\gamma$ is normalized to the cavity length and the saturation power the Kerr constant, $\varepsilon = \tau_R/\tau_{Er}$ is the ratio of the round trip time $\tau_R$ to the lifetime of erbium ions at the first excited level $\tau_{Er}$;

$\chi_{p,s} = \left(\sigma_a^{(s,p)} + \sigma_e^{(s,p)}\right)/\sigma_a^{(s,p)}$, ($\sigma_a^{(s,p)}$ and $\sigma_e^{(s,p)}$ are absorption and emission cross-sections at the lasing (s) and pump (p) wavelengths); $\Delta$ is the detuning of the lasing wavelength with respect to the maximum of the gain spectrum (normalized to the gain spectral width); $d_1 = \chi_s/(1+\Delta^2)$.

Eqs. (SE 1) have been derived under approximation that the dipole moments of the absorption and emission transitions for erbium-doped silica are located in the plane that is orthogonal to the direction of the light propagation [26-28]. This results in the angular distribution of the excited ions $n(\theta)$, which can be expanded into a Fourier series as follows:

$$n(\theta) = \frac{n_0}{2} + \sum_{k=1}^{\infty} n_{1k} \cos(k\theta) + \sum_{k=1}^{\infty} n_{2k} \sin(k\theta),$$

$$f_1 = \left(\chi \frac{n_0}{2} - 1\right) + \chi \frac{n_{12}}{2}, \quad f_2 = \left(\chi \frac{n_0}{2} - 1\right) - \chi \frac{n_{12}}{2}, \quad f_3 = \chi \frac{n_{22}}{2}. \tag{SE 2}$$

Eqs. (SE 2) are derived for the case when the dipole moments of the absorption and emission transitions for Er-doped silica are located in the plane orthogonal to the direction of the light propagation [26-28].

To obtain results shown in Figs. 6-9, we used the following parameters: $\Delta = 0.025, \varepsilon = 0.22 \cdot 10^{-5}, \alpha_1 = 12.9, \alpha_2 = 2.3, \chi_s = 2.3, \chi_p = 1$ (pump at 980 nm wavelength [26-28]), $\gamma = 2 \cdot 10^{-6}$; Fig. 6: $\sigma = 0.9$ ($\xi = 0.11$), $\beta_L = 2\pi \cdot 0.001, \beta_c = 0, I_p = 55$; Fig.7: $\sigma = 0.95$ ($\xi = 0.05$), $\beta_L = 0, \beta_c = 2\pi \cdot 0.001, I_p = 55$; Fig. 8: $\sigma = 0.8$ ($\xi = 0.22$), $\beta_L = 0, \beta_c = 2\pi \cdot 0.001, I_p = 55$; Fig. 9: $\sigma = 0.8$ ($\xi = 0.22$), $\beta_L = 0, \beta_c = 0, I_p = 29$.